\documentclass[useAMS,usenatbib]{mn2e}
\usepackage[]{graphicx}
\usepackage{times}
\usepackage{color}

\title[I. Quasi-periodic Optical and Supersoft Variability]
       {Optical and X-ray Properties of CAL~83: I. Quasi-periodic Optical and Supersoft Variability}

\author[A. F. Rajoelimanana et al.]{A. F. Rajoelimanana$^{1,2}$\thanks{E-mail: andry@saao.ac.za}, P. A. Charles$^{3}$, P.J. Meintjes$^{4}$, A. Odendaal$^{4}$, and A. Udalski$^{5}$\\
$^{1}$South African Astronomical Observatory, P.O. Box 9, Observatory, 7935, South Africa \\
$^{2}$Astrophysics, Cosmology and Gravity Centre (ACGC), University of Cape Town, Private Bag X3, Rondebosch, 7701, South Africa \\
$^{3}$School of Physics and Astronomy, Southampton University, Southampton SO17 1BJ \\
$^{4}$Department of Physics, University of the Free State, South Africa\\
$^{5}$Warsaw University Observatory, Aleje Ujazdowskie 4, 00-478 Warsaw, Poland}

\begin{document}

\date{12 April 2013, accepted for publication in MNRAS}

\pagerange{\pageref{firstpage}--\pageref{lastpage}}\pubyear{2013}

\maketitle

\label{firstpage}

\begin{abstract}

We have studied the long-term ($\sim$ years) temporal variability of the prototype supersoft X-ray source (SSS) CAL~83 in the LMC, using data from the MACHO and OGLE projects. The CAL~83 light curve exhibits dramatic brightness changes of $\sim$~1~mag on timescales of $\sim$~450~days, and spends typically $\sim$~200~days in the optical low state. Combined with archival {\it XMM-Newton} X-ray observations these represent the most extensive X-ray/optical study to date of this system, and reveal in much greater detail that the X-ray light curve is anti-correlated with the optical behaviour. This is remarkably similar to the behaviour of the ``transient'' SSS, RX~J0513.9-6951, where the SSS outbursts recur on a timescale of $\sim$~168~days, and also anti-correlate with the optical flux.  We performed simple blackbody fits to both high and low state X-ray spectra, and find that the blackbody temperature and luminosity decrease when the optical counterpart brightens.  We interpret these long-term variations in terms of the limit-cycle model of \cite{HachisuKato03a}, which provides further support for these systems containing massive ($\sim~1.3~\mathrm{M_{\sun}}$) white dwarfs. In addition, we have refined their orbital periods in the MACHO and OGLE-III light curves to values of 1.047529(1)~days and 0.762956(5)~days for CAL~83 and RX~J0513.9-6951, respectively.

\end{abstract}

\section{Introduction}
\label{intro}

Luminous SSS are a class of X-ray object first observed with the {\it Einstein} X-ray Observatory in the late 1970s \citep{long81}, of which CAL~83 and CAL~87 were the prototypes. Subsequently enhanced in number by the {\it ROSAT} soft X-ray surveys in the 1990s (\citealt{kahabka06}, for a thorough review, most recently updated by \citealt{Kato10}), their defining characteristic is their extreme luminosity ( up to $L_{x}\sim~10^{39}~\mathrm{erg~s^{-1}}$) at very soft ($<0.5$~keV) X-ray energies. As such, the SSS are rendered almost undetectable in the Galactic Plane due to interstellar extinction. Consequently, the known population of SSS are predominantly in the Magellanic Clouds and M31, although short SSS phases have been observed in a number of classical and recurrent novae in the Galaxy (see e.g. Bode (2011,2012)).

While it was initially suggested that the SSS contained Black Hole accretors, the combination of high luminosity and low temperature led \citet[][hereafter vdH92]{vdH92} to propose that SSS were actually white dwarfs accreting at or close to the Eddington Limit.  Such luminosities can be sustained via continuous nuclear burning of accreted material  on the surface of the white dwarf.  But in order to achieve this a much higher mass transfer rate is required than is observed in most cataclysmic variables. This in turn requires that the donor have a mass comparable to or greater than its accreting white dwarf, which will lead to thermally unstable mass transfer \citep[e.g.][]{paczynski71}.

With SSS phases now exhibited in many classical and recurrent nova outbursts \citep{bode10}, the vdH92 accreting white dwarf model has become a well-established paradigm. However, confirmation of the inverted mass ratio and very high mass transfer rates in order to explain the very highest luminosity SSS rests on two independent, but still circumstantial, pieces of observational evidence:

\begin{itemize}
\item \citet{southwell96} observed bipolar emission line components in the transient SSS, RX~J0513.9-6951, whose velocities were comparable to the escape velocity of a white dwarf.

\item \citet{mcgowan05} used {\it XMM-Newton} to observe RX~J0513.9-6951 through one of its transient SSS events. These revealed that, as the SSS component fades, the effective radiating radius {\it increases}, shifting the peak of the emission to longer wavelengths and hence fading from view. This supported the model of \citet{southwell96} who had noted that the SSS component occurs when the optical brightness {\it decreases}, and proposed that this is due to a {\it contraction} in the size of the white dwarf.
\end{itemize}

\citet{greiner02} have discussed several mechanisms, such as photospheric adjustments, cessation of nuclear burning, absorption of X-ray emission by material above the white dwarf surface, amongst others, to explain the X-ray off-states seen in CAL~83. They studied the early MACHO data of CAL~83, which suggested that there were three different optical states (low, intermediate, and high). Unfortunately, only two X-ray off-states had been detected during that time, in 1996 \citep{kahabka96} and 1999 \citep{greiner02}, and the variations in the MACHO light curve are complex compared to the OGLE-III lightcurve. They concluded that the most likely explanation of the X-ray variations is the photospheric expansion/contraction model of \citet{southwell96}.

In an alternative model, \citet{HachisuKato03a}, have used the more regular, recurrent optical high and low states of RX~J0513.9-6951 to develop a ``limit cycle'' model, in which the SSS phase drives a powerful wind from the disc.  This wind is capable of driving material from the donor so as to switch off effectively the mass transfer, thereby bringing the SSS phase to an end.  The donor then recovers, mass transfer resumes, the SSS switches on again, and the whole cycle continuously repeats.  Further studies by \citet{burwitz08} showed that the duration of the optical high and low states, which they called ``cycle-length'', varies for each cycle. They suggested these variations to be caused by changes in the mass transfer rate (by a factor of $\sim$~5) on timescales of years.

Extensive X-ray observations of nearby galaxies, especially M31, have been made with {\it ROSAT} \citep{kahabka99}, {\it Chandra} \citep{distefano03,distefano04} and {\it XMM-Newton} \citep{pietsch05}, revealing a large number of extra-galactic SSS.  These studies have been combined in a multi-wavelength (X-ray, UV, optical) census of SSS in M31 by \citet{orio10}.

Most of SSSs discovered so far are in distant galaxies such as in M101 \citep{pence01}, M81 \citep{swartz03}, NGC 4697 \citep{sarazin01}. Several of these SSSs have been found to have very high luminosities ($L_{\mathrm{x}}\sim 10^{39}-10^{41}~\mathrm{erg~s^{-1}}$), and it has been suggested that they may form a new class of SSS called ``ultra-luminous supersoft X-ray sources" (ULSs), see e.g. \citet{distefano10}. Their luminosities are extremely super-Eddington for a normal white dwarf, moreover these sources exhibit state changes, they have a supersoft blackbody spectrum during the outburst and a hard power-law tail in the low state ($L_{\mathrm{x}}\sim 10^{36}~\mathrm{erg~s^{-1}}$). This make ULS good candidates to be intermediate-mass black holes (IMBHs) \citep{kong05}. The prototype of this class is M101-ULX1, which exhibited a supersoft spectrum with blackbody temperature of $50-100$~eV, and bolometric luminosity of about $10^{41}~\mathrm{erg~s^{-1}}$ during its 2004 outburst \citep{kong04}. However, this interpretation is still highly controversial, and the use of real WD atmosphere models (rather than simple black-body spectra) has already reduced the apparent luminosity in a number of cases \citep{li12}.

In this paper, in order to provide additional observational constraints on the various models proposed for SSS properties, we have studied the long-term optical variations of the LMC SSS CAL~83 and RX~J0513.9-6951 using their combined light curves from the MACHO and OGLE project databases.  We compare their variability properties, with particular emphasis on the optical/X-ray correlated behaviour. We accomplish this by using archival X-ray data of CAL~83 obtained by {\it XMM-Newton}.

\section{Observations and results}

\subsection{MACHO and OGLE optical light curves}

CAL~83 has been monitored by the MACHO and OGLE projects since 1993, which now provide us with a combined light curve that spans 18~years. The MACHO observations were made with the 1.27~m telescope located at Mount Stromlo Observatory, Australia \citep{alcock96}.  This provides photometry in two passbands, a red band which we refer to as $\mathcal{R}$-band ($\sim$ 6300-7600{\AA}, a slightly longer effective wavelength than Johnson R) and a blue band we refer to as $\mathcal{V}$-band ($\sim$ 4500-6300\AA, slightly shorter effective wavelength than Johnson V). The data were taken during the interval 1993 to 2000 and are available through the MACHO website\footnote{wwwmacho.anu.edu.au}. The data are in instrumental magnitudes.

The OGLE project has used the 1.3~m  Warsaw telescope at Las Campanas Observatory, Chile, with Phase III running from 2001 until 3 May 2009. The photometric data for the OGLE projects are taken in the I-band. OGLE-III data for optical counterparts of X-ray sources located in the fields observed regularly by the OGLE-III survey are available through the X-ray variables OGLE monitoring (XROM) website\footnote{ogle.astrouw.edu.pl/ogle3/xrom/xrom.html} \citep{udalski08}.

The OGLE-III data were calibrated using observations of \citet{landolt92} standard stars which overlap with the OGLE-II fields \citep{udalski08a}. These observations were then used to determine the offset between the OGLE-III instrumental magnitudes and the standard system, which we subsequently used to calibrate the remaining OGLE-III fields. Unfortunately, both CAL~83 and  RX~J0513.9-6951 are not covered by the OGLE-II project, therefore we do not have any overlap between MACHO and OGLE data (usually from 1997 to 2000).

The light curves were then analysed using the Starlink PERIOD\footnote{www.starlink.rl.ac.uk/star/docs/sun167.htx/sun167.html} time-series analysis package, which is designed to search for periodicities in unevenly sampled data. It has a variety of temporal analysis options, including the Lomb-Scargle periodogram \citep[LS periodogram,][]{lomb76,scargle82}, as well as the capability to fold any data with a given period.

\subsubsection{Long-term variations}

\begin{figure*}
\scalebox{1.12}{\includegraphics{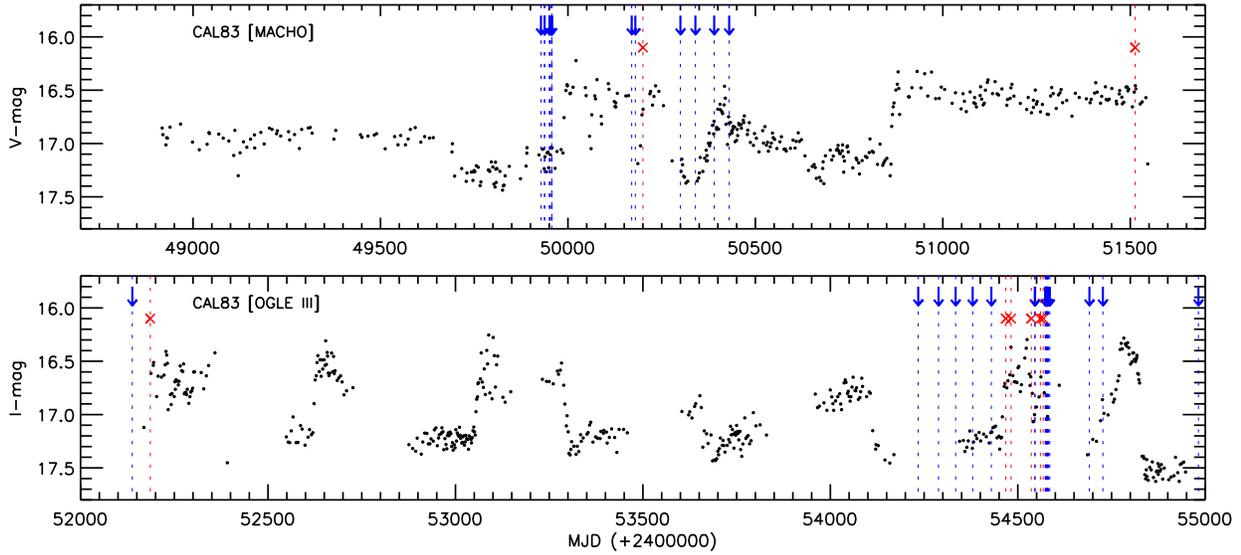}}
\caption{MACHO (top) and OGLE-III (bottom) light curves of CAL~83, showing a regular dimming of $\sim$ 1~mag every 450~days, which lasts for $\sim$~200~days. The arrows and crosses are at times of X-ray observations, and indicate X-ray on- and off-states, respectively. This clearly demonstrates that the X-ray off-states occur only during optical high states.}
\label{cal83long} 
\end{figure*}

The MACHO and OGLE projects both included many regularly monitored fields within the Magellanic Clouds, thereby providing systematic monitoring of a significant number of luminous X-ray sources. Figure \ref{cal83long} shows the combined MACHO and OGLE light curve of CAL~83. The MACHO light curve exhibits irregular variability, showing a long-lived intermediate state for the first $\sim$~700~days, and similar duration high state towards the end of the observations.  However, the OGLE-III light curve is rather different, exhibiting a regular succession of optical high and low states on timescales of 450~days with a difference in brightness of $\sim$ 1~mag.  CAL~83 spends roughly the same time ($\sim$ 200~days) within each state, which is much longer than the duration of the transitions between states (typically $\sim ~5-10$~days).

\begin{figure*}
\scalebox{1.12}{\includegraphics{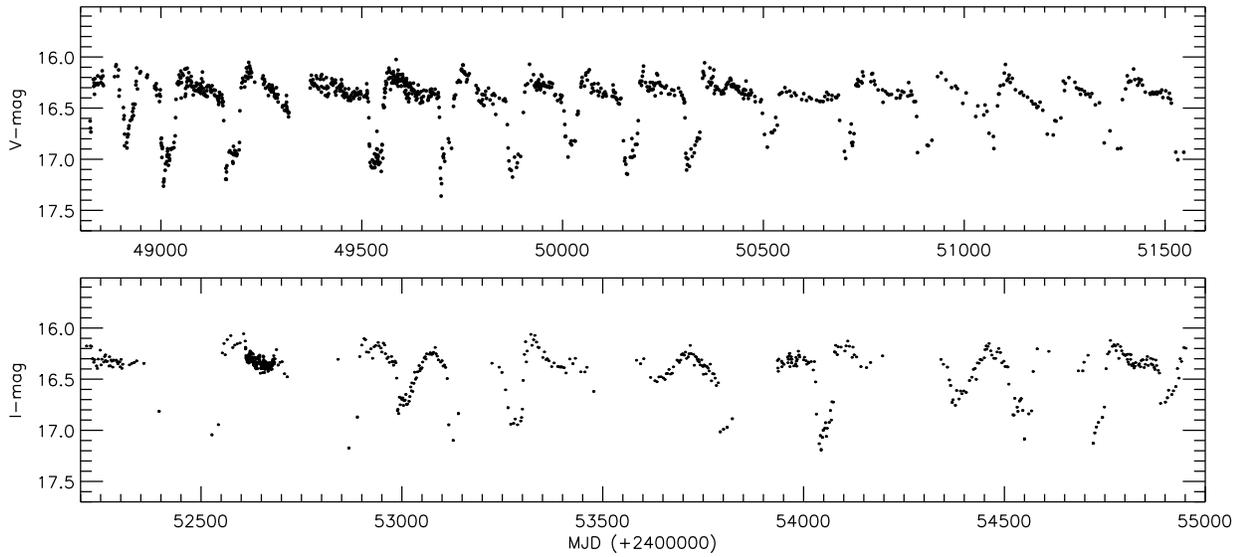}}
\caption{MACHO (top) and OGLE-III (bottom) light curves of RX~J0513.9-6951, showing a regular dimming of $\sim$~1~mag every 168~days, which lasts for $\sim$~30~days.}
\label{rxj0513long} 
\end{figure*}

\begin{figure}
\scalebox{0.53}{\includegraphics{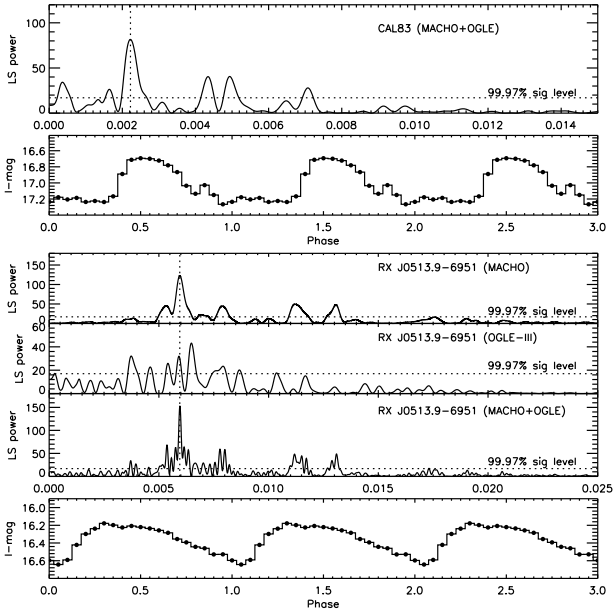}}
\caption{Lomb-Scargle periodogram of CAL~83 and its folded light curve (from Figure~\ref{cal83long}) on the 450~d period (top) and Lomb-scargle periodogram of the independent MACHO and OGLE-III as well as the combined  data of RX~J0513.9-6951, and the combined light curves folded on the superorbital period of 168~d (bottom).}
\label{long_period_all} 
\end{figure}

It is instructive to compare the properties of the LMC supersoft source RX J0513.9-6951 with CAL~83.  Its MACHO and OGLE light curves are shown in Figure \ref{rxj0513long}. RX J0513.9-6951 exhibits similar behaviour to CAL~83, but with a regular dimming on a shorter timescale ($\sim$~168~days), and it also spends less time in the optical minimum state.  These timescales are nicely revealed in the Lomb-Scargle periodograms (Figure \ref{long_period_all}) of the combined MACHO and OGLE light curves, which show broad peaks at periods of P = 450$\pm$3~days and 168$\pm$1~days for CAL~83 and RX J0513.9-6951, respectively.  We then folded the data on these long periods, with the results shown in Figure \ref{long_period_all}.

More importantly, for both systems, the regular dimming in the optical is associated with the turn-on of the supersoft source component.  This will be discussed further in the later sections.

\subsubsection{Orbital modulations}

We also exploit the OGLE-III data to search for short-term variations that could be related to the orbital properties of these systems. The light curves were detrended by subtracting a linear fit to the optical low and high states. The detrended light curves were then analysed using PERIOD.

\begin{figure}
\scalebox{0.56}{\includegraphics{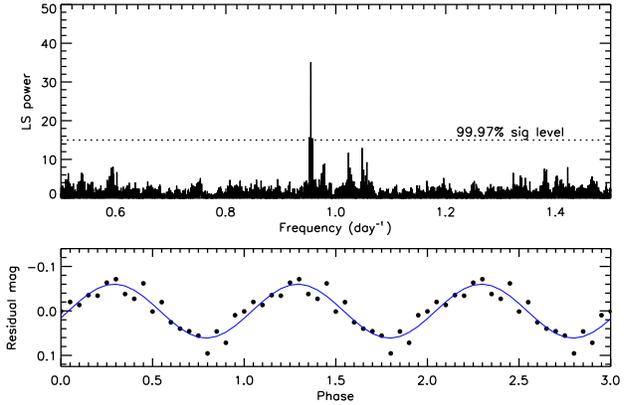}}
\caption{Lomb-Scargle periodogram of CAL~83 (top) showing a very significant peak at the presumed $P_{{orb}}$ of 1.047529~days. Bottom: Light curve folded on this period and fitted with a sinusoid (Blue line).}
\label{cal83_per} 
\end{figure}

The power spectrum of CAL~83 (figure~\ref{cal83_per}) shows a significant peak at 1.047529~$\pm$~0.000001~days, which is very close to the modulation reported by \citet{smale88} of $P_{orb}$ = 1.0436~days based on photometric and spectroscopic data. The folded light curve on the presumed orbital period of $P_{{orb}}$ = 1.047529~days is shown in Figure~\ref{cal83_per} (bottom). It is sinusoidal, with a semi-amplitude of $\sim$~0.1~mag.

\begin{figure}
\scalebox{0.56}{\includegraphics{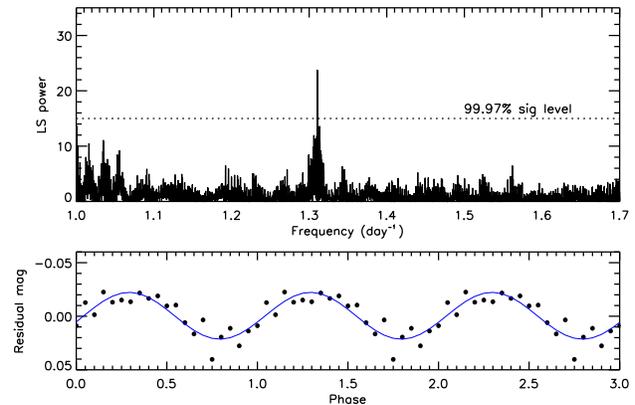}}
\caption{Lomb-Scargle periodogram of RX J0513.9-6951 (top) showing the peak at the presumed $P_{{orb}}$ of 0.762956~days. Bottom: Light curve folded on this period and fitted with a sinusoid (Blue line).}
\label{rxj_per} 
\end{figure}

We performed the same analysis for RX J0513.9-6951. The periodogram of its detrended OGLE-III light curve reveals a significant peak at 0.762956$\pm$0.000005~days (figure~\ref{rxj_per}), again very close to previously determined values found in the MACHO data alone by \citet{alcock96} and \citet{cowley02}, and spectroscopically by \citet{southwell96}. The light curve (folded on $P_{orb}$ = 0.762956~days) is shown in Figure~\ref{rxj_per}, and is also well fitted by a sinusoidal modulation of semi-amplitude $\sim$~0.2~mag.

\subsection{X-ray observations of CAL~83 from XMM.}

CAL~83 was observed by {\it XMM-Newton} (proposal ID 050086 and 050653) between May 2007 and May 2009. The observation logs of the X-ray measurements are summarised in Table \ref{obslogs}. The observations were obtained with the EPIC-PN \citep{struder01}, EPIC-MOS \citep{turner01} and RGS instruments. The data reductions were performed using SAS version 11.0.0. There are another four available {\it Chandra} observations of CAL~83 taken in 1999 and 2001, but unfortunately three of them are during the X-ray off state, and the X-ray on state has already been reported by \citet{lanz05}. They are included in Figure \ref{cal83long} and are consistent with the anti-correlation.

\begin{table}
\centering
\caption{{\it XMM-Newton} EPIC observations of CAL~83 between May 2007 and May 2009.}
\scriptsize{
\begin{tabular}{llrcc}\hline\hline
\multicolumn{1}{c}{\bf{Observation}}&
\multicolumn{1}{c}{\bf {EPIC}}&
\multicolumn{3}{l}{\bf {~~~~~~~~Start~~~~~~~~~~Exposure~~~~X-ray}}\\

\multicolumn{3}{l}{\bf {~~~~~ID~~~~~~~~~Instrument~~~~~~time(UT)}}&
\multicolumn{1}{c}{\bf {time(s)}}&
\multicolumn{1}{c}{\bf {State}} \\
\hline
0500860201 	& PN		& 2007-05-13 22:09:25 	 	& 11464	&\\
 		& M1/M2 	& 	     22:03:52 	 	& 11657	&\\
0500860301 	& PN		& 2007-07-06 23:37:53 	 	& 10464	&\\
		& M1/M2 	& 	     23:32:20 	 	& 10657	&\\
0500860401 	& PN		& 2007-08-21 15:17:17 	 	& 7464	&\\
		& M1/M2 	& 	     15:11:44 	 	& 7657	&\\
0500860501 	& PN		& 2007-10-05 01:13:07 	 	& 12507	&\\
		& M1/M2 	&	     23:49:45 	 	& 14072	&\\
0500860601 	& PN		& 2007-11-24 21:07:07 	 	& 20031	&\\
		& M1/M2 	&	     21:01:34 	 	& 19553	&\\
0500860701 	& PN		& 2008-01-16 13:30:30 	 	& 10464	& off\\
		& M1/M2 	&	     13:24:57 	 	& 10657	&\\
0500860801 	& PN		& 2008-03-10 10:20:35 	 	& 6464	& off\\
		& M1/M2 	&	     10:15:02 	 	& 6657	&\\
0500860901 	& PN		& 2008-03-20 00:39:20 	 	& 7264	&\\
		& M1/M2 	&	     00:33:47 	 	& 7457	&\\
0506530201 	& PN		& 2008-04-03 18:46:54 	 	& 14464	&weak\\
		& M1/M2 	&	     18:41:21 	 	& 14657	&\\
0506530301 	& PN		& 2008-04-11 06:08:21 	 	& 5464	& off\\
		& M1/M2 	&	     06:02:48 	 	& 5657	&\\
0506530401 	& PN		& 2008-04-16 14:15:17 	 	& 4595	& off\\
		& M1/M2 	&	     14:06:43 	 	& 4936	&\\
0506530501 	& PN		& 2008-04-17 13:46:11 	 	& 10764	&\\
		& M1/M2 	&	     13:40:38 	 	& 10957	&\\
0506530601 	& PN		& 2008-04-19 06:49:33 	 	& 5464	&\\
		& M1/M2 	&	     06:44:00 	 	& 5657	&\\
0506530801 	& PN		& 2008-04-20 22:44:41 	 	& 11164	&weak\\
		& M1/M2 	&	     22:39:08 	 	& 11357	&\\
0506530901 	& PN		& 2008-04-21 02:16:21 	 	& 12264	&\\
		& M1/M2 	&	     02:10:48 	 	& 12457	&\\
0506531001 	& PN		& 2008-04-21 18:53:55 	 	& 8668	&\\
		& M1/M2 	&	     18:48:22 	 	& 8967	&\\
0506531201 	& PN		& 2008-04-23 11:26:20 	 	& 6964	& weak\\
		& M1/M2 	&	     11:20:47 	 	& 7157	&\\
0506531301 	& PN		& 2008-04-25 08:19:28 	 	& 9164	&\\
		& M1/M2 	&	     08:13:55 	 	& 9357	&\\
0506531401 	& PN		& 2008-04-29 00:46:22 	 	& 13664	&\\
		& M1/M2 	&	     00:40:49 	 	& 13857	&\\
0506531501 	& PN		& 2008-08-12 14:56:26 	 	& 6464	&\\
		& M1/M2 	&	     14:50:53 	 	& 6657	&\\
0506531601 	& PN		& 2008-09-17 11:16:18 	 	& 6364	& weak\\
		& M1/M2 	&	     11:10:45 	 	& 6557	&\\
0506531701 	& PN		& 2009-05-30 08:06:47 	 	& 45664	&\\
		& M1/M2 	&            08:01:14 	 	& 45857	&\\
\hline

\label{obslogs}
\end{tabular}

}
\end{table}

\subsubsection{XMM EPIC X-ray light curve}

\begin{figure*}
\scalebox{0.82}{\includegraphics{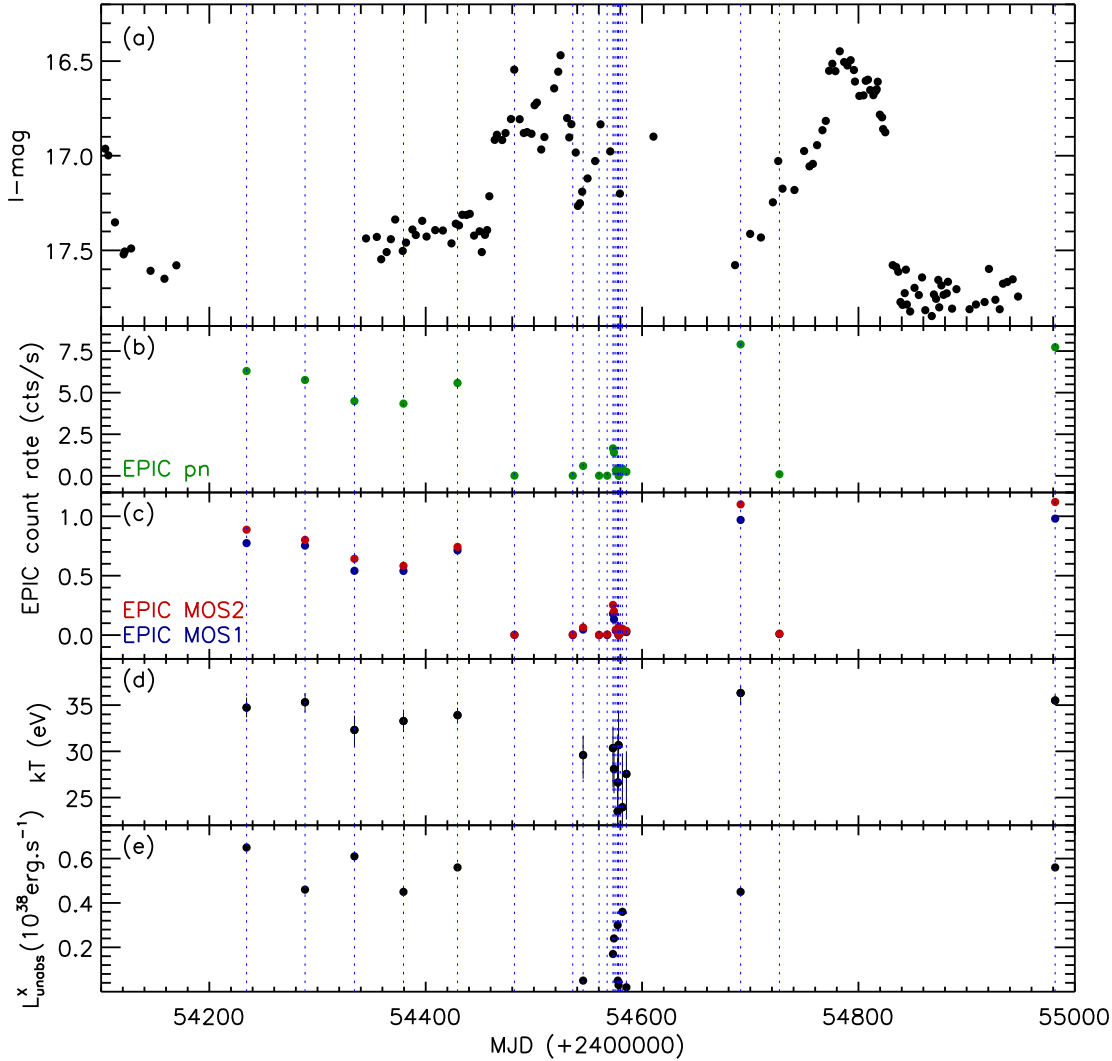}}
\caption{OGLE-III (a) and {\it XMM-Newton}/EPIC PN (b), MOS1, and MOS2 (c) light curves of CAL~83, which shows a clear anti-correlation between the optical and X-ray states. Times of the {\it XMM-Newton} observations are indicated by dashed vertical lines on the optical light curve. Panels (d) and (e) show the temperature and luminosity evolution, respectively, based on the X-ray spectral fitting (see text).}

\label{opt_xray} 
\end{figure*}

We extracted the EPIC-PN and EPIC-MOS light curves in the energy band $0.2-1.0$~keV. The mean EPIC-PN and EPIC/MOS1-MOS2 count rates from each observation are plotted with the OGLE-III I-band light curve in figure \ref{opt_xray}.

The X-ray mean count rates are highly variable and show two different X-ray states: a high state where the EPIC count rates are at a maximum ($\ge 4.5~\mathrm{cts~s^{-1}}$ for EPIC-PN, and $\ge 0.55~\mathrm{cts~s^{-1}}$ for EPIC-MOS), and a low state where the X-ray emission from the source is weak or completely off (and which occur during optical maxima). The anti-correlation between the optical and X-ray fluxes is clearly seen in figure \ref{opt_xray}. Unfortunately,  we have only a few (4) optical data points between MJD $54555 - 54590$ where most of the X-ray observations took place. This makes it impossible for us to distinguish whether the X-ray variations lag or lead those in the optical.

\begin{figure*}
\scalebox{0.4}{\includegraphics{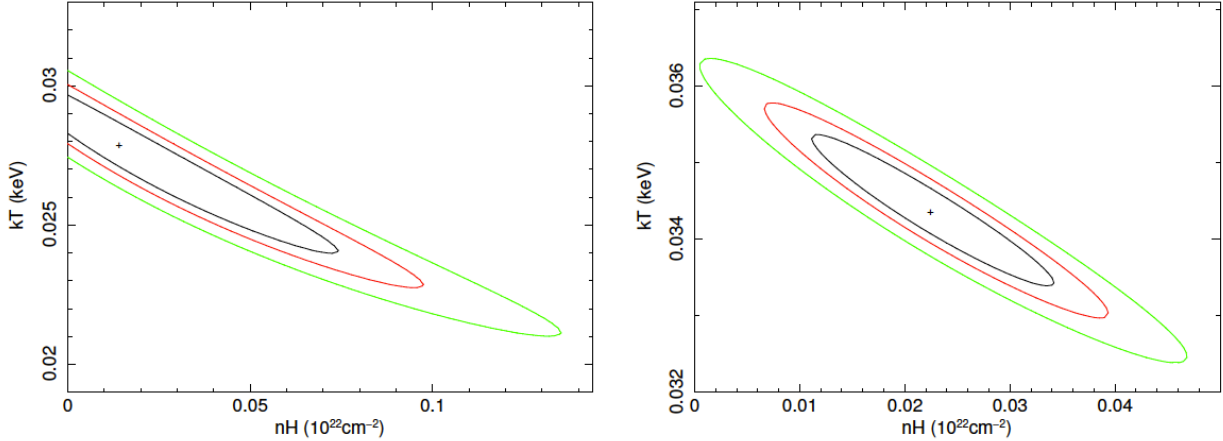}}
\caption{Confidence contours in ($N_{\mathrm{H}} - kT$) plane for optical high (left) and optical low (right) state. The contours are plotted at 1 $\sigma$ (68\%), 90\%, and 99.0\% confidence levels for two parameters.}

\label{chi_map} 
\end{figure*}

\subsubsection{Low resolution X-ray spectra}

The X-ray spectral analysis was done using the latest XSPEC version (v 12.7.0), and the spectra were extracted using only photons in the energy range $0.3-1.0$~keV. We used a simple blackbody model ({\it bbody}) together with the T\"{u}bingen-Boulder ISM absorption ({\it TBabs}) model, ISM abundances from \citet{Wilms00}, and the photoelectric absorption cross-sections from \citet{Balucinska92}. The column density was fixed to the value derived by \citet{gansicke98} of $N_{\mathrm{H}}$ = $6.5 \times 10^{20}~\mathrm{cm^{-2}}$ from {\it Hubble Space Telescope} UV observations of CAL~83. We add an additional absorption component with free column density and abundances set to 0.5 solar (LMC abundances) for elements heavier than helium.

The results from a simple blackbody model fit ({\it bbody}) to the EPIC-PN spectra are summarised in table~\ref{res}. The values of the blackbody temperature, $kT$, inferred X-ray luminosity (assuming an LMC distance of 50 kpc, \citep{feast99}), $L_{BB}$, and white dwarf radius, $R_{WD}$, vary significantly with the optical state of the source. The blackbody temperature has a mean value of $kT \sim$ 35 eV during optical low states when the X-ray flux is highest, and reduces to $\le 30$ eV during optical high states. Figure~\ref{opt_xray} (bottom) shows the evolution of $kT$ and $L^{{x}}_{{unabs}}$ with time, where they are clearly anti-correlated with its optical brightness. The $\chi^2$ contour maps of the optical low state spectra (OBSID: 0500860201) and four merged optical high state spectra (OBSID: 0506530901, 0506531001, 0506531301, 0506531401) are presented in Figure \ref{chi_map}, showing how the column density and the blackbody temperature correlate with each other.

\begin{table}
\centering
\caption{X-ray Off-state Observations}

\begin{tabular}{lcc}
\hline
  \multicolumn{1}{c}{X-ray}	&
  \multicolumn{1}{c}{Obs}	&
  \multicolumn{1}{c}{References} \\

  \multicolumn{1}{c}{Mission}	&
  \multicolumn{1}{c}{date}	&
  \multicolumn{1}{c}{}\\
\hline
{\it ROSAT} 	& 1996 Apr 28	& \citet{kahabka96} \\
{\it Chandra}	& 1999 Nov 30	& Greiner \& Di Stefano (2002) \\
{\it Chandra}	& 2001 Oct 03	& \citet{lanz05} \\
{\it Swift}	& 2008 Jan 02	& \citet{greiner08} \\
{\it XMM-Newton}& 2008 Jan 16   & \\
{\it XMM-Newton}& 2008 Mar 10	& \\
{\it XMM-Newton}& 2008 Apr 11	& \\
{\it XMM-Newton}& 2008 Apr 16	& \\ 
\hline

\label{offstate}
\end{tabular}
\end{table}

\begin{figure*}
\scalebox{0.98}{\includegraphics{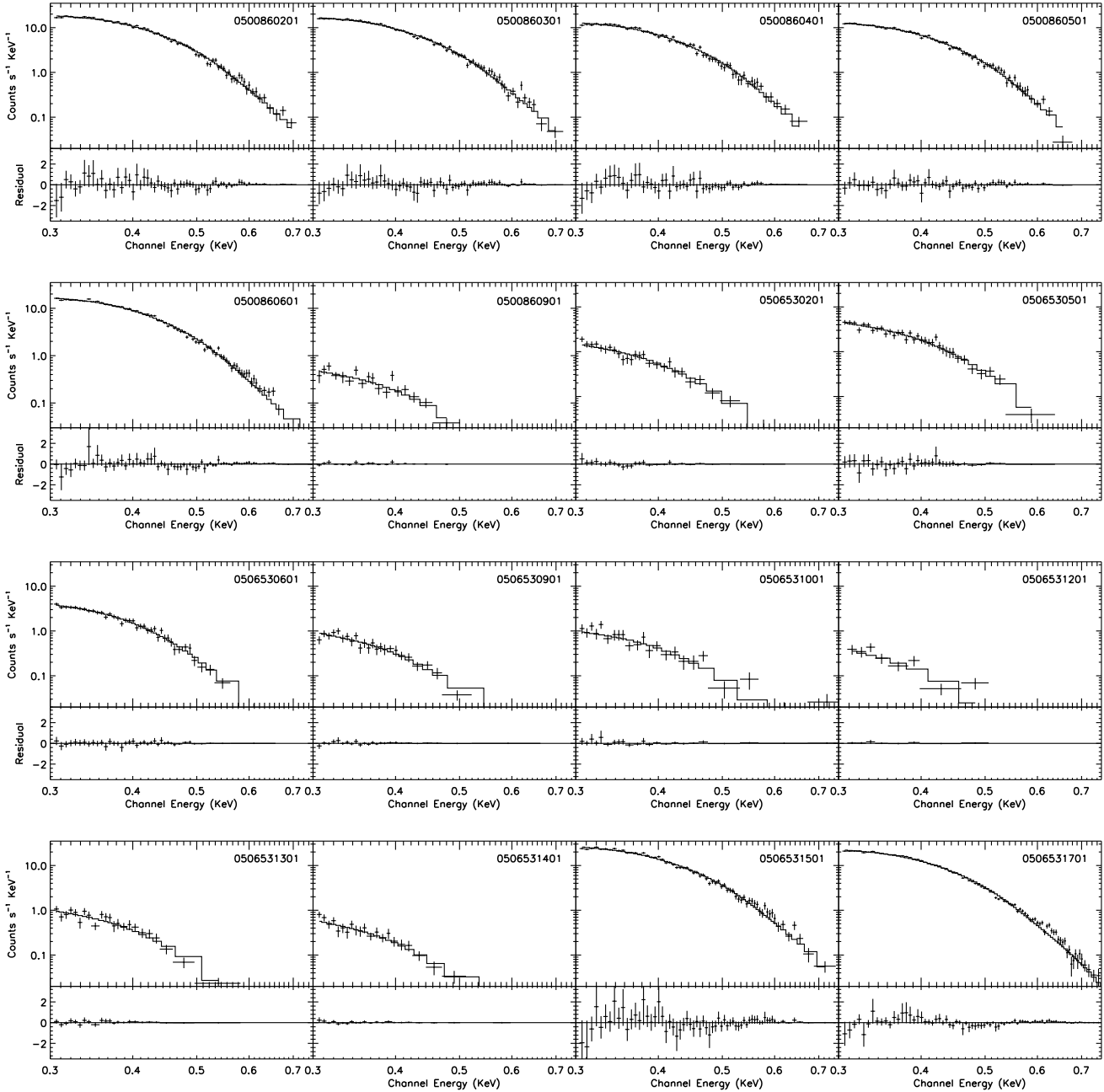}}
\caption{EPIC/PN spectra of CAL~83 plotted with the best blackbody fits, using a fixed column density of {\it $N_{\mathrm{H}}$} = 6.5$\times 10^{20}~\mathrm{cm^{-2}}$ \citep{gansicke98} and an additional absorption with free {\it$N_{\mathrm{H}}$} (and abundances set to 0.5 for elements heavier than helium) relative to the source intrinsic absorptions.}
\label{spectra_pl} 
\end{figure*}

\begin{table*}
\centering
\caption{CAL~83 {\it XMM-Newton} Spectral Fitting Results.}

\begin{tabular}{|r|c|c|c|c|c|c|c|c|}
\hline\noalign{\smallskip}
  \multicolumn{1}{|c|}{OBSID}	&
  \multicolumn{1}{|c|}{Additional $N_H$}	&
  \multicolumn{1}{c|}{$kT$}	&
  \multicolumn{1}{c|}{$L_{BB}$}	&
  \multicolumn{1}{c|}{$R_{\mathrm{WD}}$}	&
  \multicolumn{1}{c|}{$L_{\mathrm{X}}$}	&
  \multicolumn{1}{c|}{$L_{\mathrm{unabs}}$}	&
  \multicolumn{1}{c|}{$\chi^2_\nu$}&
  \multicolumn{1}{c|}{DOF} \\

  \multicolumn{1}{|c|}{}	&
  \multicolumn{1}{|c|}{$\mathrm{10^{20}cm^{-2}}$}	&
  \multicolumn{1}{c|}{(eV)}	&
  \multicolumn{1}{c|}{\textrm{10$^{38}$erg~s$^{-1}$}}	&
  \multicolumn{1}{c|}{$\mathrm{10^{9}cm}$}	&
  \multicolumn{1}{c|}{$\mathrm{10^{36}erg~s^{-1}}$}	&
  \multicolumn{1}{c|}{$\mathrm{10^{38}erg~s^{-1}}$}	&
  \multicolumn{1}{c|}{}	&
  \multicolumn{1}{c|}{} \\
\hline

\rule{0pt}{3ex} 0500860201   &   2.0$_{-1.2}^{+1.0}$    &   34.7$_{-1.1}^{+1.1}$   &   4.0$_{-1.2}^{+1.6}$   &   4.61   &   3.59   &   0.65   &     1.21   &     63\\
\rule{0pt}{3ex} 0500860301   &   0.8$_{-0.8}^{+1.1}$    &   35.3$_{-1.0}^{+1.0}$   &   2.7$_{-0.6}^{+1.1}$   &   3.66   &   3.51   &   0.46   &     1.02   &     61\\
\rule{0pt}{3ex} 0500860401   &   2.6$_{-1.8}^{+2.0}$    &   32.3$_{-1.9}^{+1.6}$   &   4.9$_{-2.2}^{+5.1}$   &   5.88   &   2.52   &   0.61   &     1.09   &     53\\
\rule{0pt}{3ex} 0500860501   &   1.4$_{-1.3}^{+1.4}$    &   33.3$_{-1.2}^{+1.2}$   &   3.3$_{-1.1}^{+1.9}$   &   4.54   &   2.65   &   0.45   &     1.16   &     56\\
\rule{0pt}{3ex} 0500860601   &   1.4$_{-0.9}^{+0.8}$    &   33.9$_{-0.9}^{+0.8}$   &   3.8$_{-0.9}^{+1.3}$   &   4.69   &   3.37   &   0.56   &     1.68   &     65\\
\rule{0pt}{3ex} 0500860901   &   3.6$_{-3.6}^{+13.3}$    &   26.6$_{-9.6}^{+5.2}$   &   0.9$_{-0.9}^{+1.3}$   &   3.80   &   0.10   &   0.05   &     1.13   &     15\\
\rule{0pt}{3ex} 0506530201   &   0.0 &   29.6$_{-2.5}^{+2.1}$   &   0.5$_{-0.1}^{+1.0}$   &   2.33   &   0.32   &   0.05   &     0.65   &     23\\
\rule{0pt}{3ex} 0506530501   &   1.6$_{-1.6}^{+6.0}$    &   30.3$_{-4.2}^{+2.3}$   &   1.7$_{-0.8}^{+1.3}$   &   3.96   &   0.77   &   0.17   &     0.75   &     32\\
\rule{0pt}{3ex} 0506530601   &   2.8$_{-2.8}^{+3.6}$    &   28.1$_{-2.6}^{+2.5}$   &   3.4$_{-1.0}^{+1.1}$   &   6.54   &   0.67   &   0.24   &     0.64   &     36\\
\rule{0pt}{3ex} 0506530901   &   8.3$_{-7.8}^{+7.9}$    &   23.5$_{-5.8}^{+5.7}$   &   10.7$_{-1.1}^{+1.1}$  &  16.45   &   0.12   &   0.30   &     0.72   &     22\\
\rule{0pt}{3ex} 0506531001   &   0.0    &   30.7$_{-4.9}^{+3.7}$   &   0.3$_{-0.1}^{+2.2}$   &   1.60   &   0.21   &   0.03   &     1.21   &     20\\
\rule{0pt}{3ex} 0506531301   &   8.9$_{-9.3}^{+8.9}$    &   24.0$_{-6.5}^{+5.8}$   &   10.5$_{-1.2}^{+1.2}$  &  15.71   &   0.13   &   0.36   &     1.03   &     20\\
\rule{0pt}{3ex} 0506531401   &   0.1$_{-0.1}^{+5.5}$    &   27.6$_{-4.9}^{+2.5}$   &   0.4$_{-0.1}^{+0.6}$   &   2.27   &   0.14   &   0.02   &     0.68   &     19\\
\rule{0pt}{3ex} 0506531501   &   0.4$_{-0.4}^{+1.1}$    &   36.3$_{-1.3}^{+0.7}$   &   2.4$_{-0.3}^{+1.2}$   &   3.28   &   4.04   &   0.45   &     1.40   &     62\\
\rule{0pt}{3ex} 0506531701   &   1.7$_{-0.6}^{+0.2}$    &   35.5$_{-0.5}^{+0.6}$   &   3.3$_{-0.5}^{+0.4}$   &   3.98   &   3.58   &   0.56   &     2.86   &     82\\

\hline
\multicolumn{9}{p{15cm}}{Blackbody fits to the EPIC/PN spectra using a fixed value of $N_{\mathrm{H}}$ = 6.5$\times 10^{20}~\mathrm{cm^{-2}}$ \citep{gansicke98} and an additional $N_{\mathrm{H}}$ with abundances set to 0.5 for elements heavier than helium relative to the source intrinsic absorptions.}
\label{res}
\end{tabular}
\end{table*}

\section{Discussion}

\subsection*{Optical variations in CAL~83}

In supersoft sources, the accretion disc is believed to be the main source of optical flux. However, these optical emissions arise in the accretion disc as a result of reprocessed supersoft X-rays from the hot white dwarf, and hence are not directly produced by accretion processes \citep{popham96}. The reprocessed luminosity greatly exceeds that of the accretion luminosity, as the energy released by nuclear burning is significantly greater than that due to accretion onto a white dwarf, i.e. the disc luminosity will be comparable to that seen in normal galactic cataclysmic variables, at least one to two orders of magnitude below that seen here, see e.g. \citet{warner95}.

The OGLE-III light curve of CAL~83 in Figure~\ref{cal83long} shows a quasi-periodic variation on a timescale of $\sim$~450~days, with two well-defined optical states (low and high). The variation of 1 mag in optical brightness must be associated with the changes in the amount of irradiation of the accretion disk, which is most likely due to an expansion/contraction of the white dwarf photospheric radius \citep{southwell96}, and/or the irradiated areas of the accretion disc itself \citep{HachisuKato03a}. The mass accretion rate required by the vdH92 model is very high, typically $\dot{M}~\sim~10^{-7}~\mathrm{M_{\odot}yr^{-1}}$, and steady nuclear burning on the surface of the white dwarf can occur. However, when the mass accretion rate exceeds the steady nuclear burning rate, the white dwarf photosphere will expand and its effective temperature will be reduced.

The duration of the optical low (presumably, X-ray on) and high states in CAL~83 are $\sim$~200~days and $\sim$~250~days, respectively. These durations are much longer than those observed in the recurrent supersoft source RX J0513.9-6951 ($\sim$~40~days and $\sim$~140~days) \citep{southwell96, alcock96, burwitz08}. This implies that the mass accretion rate ($\dot{M}_{accr}$) and the mass of the white dwarf ($M_{WD}$) in CAL~83 are slightly lower than in RX J0513.9-6951, as the duration of the X-ray on state is inversely proportion to the mass accretion rate \citep{hachisu03, burwitz08}.  The higher mass accretion rate of RX J0513.9-6951 makes its optical luminosity (I~$\sim$~16.2) higher compared with that of CAL~83 (I~$\sim$~16.7), given that they are at the same distance and both are subject to very low extinction levels.  Nevertheless, the implied masses of the white dwarfs in both systems are $\sim~1.25-1.3~\mathrm{M_{\sun}}$.

The orbital periods of 1.05~days and 0.76~days that we have confirmed in CAL~83 and RX J0513.9-6951, respectively, (on the basis of their stability over the last $\sim$ 18 years) are much longer than ``normal'' dwarf nova systems. This also supports the vdH92 model of an evolved and more massive donor as required in order for the donor to be filling its Roche lobe in such a system.

\subsection*{X-ray variations and optical/X-ray correlations}

In the vdH92 model, a steady nuclear burning of hydrogen near the white dwarf's surface is proposed to be the main source of supersoft X-rays. This implies that variations in the X-ray flux may be caused by changes in burning rate, mass accretion rate or photospheric radius. \citet{greiner02} discussed these mechanisms in order to explain the observed X-ray off-states and optical variability in CAL~83. They conclude that the model of \citet{southwell96} (expansion/contraction of WD's photosphere) accounts for the X-ray and and optical variations.

The X-ray off-state of CAL~83 has been detected 8 times since its discovery (see table~\ref{offstate}), and all of them occur during optical high states (Figure~\ref{cal83long}). The observed X-ray spectrum slightly deviates from a blackbody fit (especially at X-ray high state). However, we can still see broad changes in its blackbody temperature and luminosity as a function of its optical brightness (Figure~\ref{opt_xray}).

The mass accretion rates in SSS are comparable or slightly higher than the nuclear burning rate. A modest changes in accretion rate will lead to an expansion of the white dwarf photosphere which enhances the irradiation of the accretion disc and the mass flow through the disc. Correspondingly, the effective temperature will drop, as the peak of the emission is shifted from supersoft X-rays towards longer wavelengths, principally into the extreme-ultraviolet. However, the source will be brighter in the optical due to the increasing contribution from the expanding (and still Eddington-limited) white dwarf emission, together with reprocessed radiation from the accretion disc/disc wind. The results from our simple blackbody fits to the spectra show a higher effective temperature ($kT_{mean}\sim$~35~eV) during optical low state, which then drops to $kT_{mean}\sim$~30~eV during optical high (if it is detectable at all). This implies that the main source of the X-ray emission becomes cooler, and the peak flux will be shifted towards longer wavelengths. This behaviour has already been seen (and quantified) in RX J0513.9-6951 by \citet{mcgowan05}.

Once the mass accretion rate drops, the  white dwarf's photosphere will contract again, thereby raising its effective temperature and the source will re-enter the supersoft phase. On the other hand, the contraction of the white dwarf's envelope will reduce the amount of disc illumination. This causes the optical brightness to reduce during the X-ray on-state. The expanding/contracting photosphere model agrees very well with our optical and X-ray observations as well as the anti-correlation between X-ray and optical fluxes. The most likely cause of the  white dwarf's radius variations is the changes in mass accretion rate. The physical mechanism underlying this process is still unclear, although several mechanisms have been suggested, such as the accretion wind evolution model \citep{hachisu03} where the strong wind from the white dwarf is colliding with and stripping off the outer layers of the slightly evolved secondary. This will attenuate the mass outflow and therefore decrease the mass transfer rate. Some authors have suggested that the reduction of the mass accretion rate is connected to the magnetic activity of the secondary star, where mass loss from the secondary is attenuated when star spots pass over the inner Lagrangian point \citep{southwell96, alcock96}. This is similar to the proposed mechanism for the drops in accretion rate in VY Scl stars \citep{livio94}, which has led to the suggestion that VY Scl stars are low-mass extensions of supersoft sources \citep{greiner10}.

In the OGLE-III light curve (Figure~\ref{cal83long}), the times of optical low and high states in CAL~83 are quite regular, apart for the last 300~days of the OGLE-III observation. Based on the expansion/contraction model the duration of the optical low state should be the duration where we expect an X-ray detection of the source. Using the analytic model for recurrent supersoft sources by \citet{kahabka95} with recurrence time $T_{rec}$~=~450~days and X-ray on time $T_{x-on}$~=~200~days (the duration of optical low state), one would expect a white dwarf mass in the range of $M_{WD}~\sim~1.32-1.38~M_{\odot}$ and mass accretion rate of $\dot{M}_{accr}\sim1.2-3.3\times10^{-7}~\mathrm{M_{\odot}yr^{-1}}$.

\section{Summary}

We have studied the optical and X-ray variations of the prototypical SSS, CAL~83. The long-term optical light curve shows a quasi-periodic variation on a timescale of 450~days with two well defined optical states. The duration of the optical low and high states (200~days and 250~days respectively) are longer than those observed in RX J0513.9-6951, which implies that the mass accretion in CAL~83 is slightly lower than in RX J0513.9-6951. Using the combined MACHO and OGLE light curves, we have refined the short-term periods of CAL~83 and RX J0513.9-6951 to values of 1.047529(1) d and 0.762956(5) d, respectively, which are consistent with previously reported values, and further strengthens their interpretation as being orbital in origin.

The results from simple blackbody fits to the EPIC/PN spectra show a difference in blackbody temperature (from $kT~\sim$~30 eV to $\sim$~35 eV) and luminosity between the optical high and low states. The optical brightness and the X-ray luminosity in CAL~83 clearly show an anti-correlation. The non-detections of X-ray emission, in eight X-ray observations since its discovery, occur only during the optical high state. This is in good agreement with the photospheric expansion/contraction model suggested by \citet{southwell96}.

\label{lastpage}

\end{document}